\documentclass[%
 aip,
cp,  
 amsmath,amssymb,
 reprint,%
]{revtex4-2}

\usepackage{graphicx}
\usepackage{dcolumn}
\usepackage{bm}

\usepackage[utf8]{inputenc}
\usepackage[T1]{fontenc}
\usepackage{mathptmx} 

\begin{document}

\title{Quantum gravitational wave function for the interior of a black hole and the generalized uncertainty principle}

\author{Dong-han Yeom}
 \email[Corresponding author: ]{innocent.yeom@gmail.com}

\affiliation{
Department of Physics Education, Pusan National University, Busan 46241, Republic of Korea \\
Research Center for Dielectric and Advanced Matter Physics, Pusan National University, Busan 46241, Republic of Korea
}


\begin{abstract}
We investigate the internal structures of a Schwarzschild black hole by solving the Wheeler-DeWitt equation. The generic bounded wave function has a bouncing point around $r \simeq M$, where $M$ is the black hole mass. Due to this quantum bouncing, there appears an ambiguity to define the arrow of time. If we introduce two arrows of time, one can then interpret that two classical spacetime is annihilated around the bouncing point. Finally, we provide a conceptual explanation based on the generalized uncertainty principle.\footnote{Proceedings of the 14th Asia-Pacific Physics Conference. Talk on November 21, 2019, Kuching, Malaysia.}
\end{abstract}

\maketitle

\section{Introduction}

The investigation of the internal structure of a black hole \cite{Hong:2008mw} is an interesting but unresolved problem. This is deeply related to several candidate theories of quantum gravity as well as their resolutions to the information loss problem.

In order to resolve the singularity inside a black hole, several proposals have been proposed. One of them is so-called the regular black hole picture \cite{AyonBeato:1999rg,Hwang:2012nn}. A black hole is regular if the black hole has no singularity. The absence of a singularity requires a violation of one of the assumptions of the singularity theorem \cite{Chen:2014jwq}. This is in principle possible but quite unnatural. If the absence of the singularity is originated by physical matters, then it seems to cause a severe inconsistency or an additional problem \cite{Brahma:2019oal}.

The next alternative is to resolve the singularity based on quantum gravitational effects. Recently, there have been several proposals about this, especially using loop quantum gravity. If the quantum gravitational effects can spread to outside the horizon \cite{Haggard:2014rza}, then these effects can reach infinity which is potentially very harmful \cite{Brahma:2018cgr}. On the other hand, if the quantum effects are limited only inside the horizon, one may ask what is the causal structure beyond the putative singularity. One possibility is to see an inner apparent horizon \cite{Bojowald:2018xxu} while the other possibility is to see a white hole region without seeing an inner horizon \cite{Ashtekar:2018lag} (see also \cite{Bouhmadi-Lopez:2019hpp}).

In this context, we may get wisdom from the canonical approach toward quantum gravity \cite{DeWitt:1967yk}. This approach means that we need to solve the Wheeler-DeWitt equation. If we can solve and understand the Wheeler-DeWitt equation, we can then provide wisdom to understand and interpret quantum gravitational effects near the singularity \cite{Bouhmadi-Lopez:2019kkt}. In addition, this will shed some light to resolve the information loss paradox.

\section{Formalism and solutions}

In order to derive the Wheeler-DeWitt equation, for simplicity, we will impose the following metric form with the spherical symmetry:
\begin{eqnarray}
ds^{2} = - N^{2}(t) dt^{2} + a^{2}(t) dR^{2} + \frac{r_{s}^{2} b^{2}(t)}{a^{2}(t)} d\Omega^{2},
\end{eqnarray}
where $N(t)$ is the lapse function, $r_{s}$ is the Schwarzschild radius, $a(t)$ and $b(t)$ are two independent canonical variables. For the classical solution of the vacuum, the following relation should be satisfied:
\begin{eqnarray}
\frac{1}{b} = a + \frac{1}{a}.
\end{eqnarray}

Using this metric ansatz, we can derive the Wheeler-DeWitt equation \cite{Cavaglia:1994yc}:
\begin{eqnarray}
\left( \frac{\partial^{2}}{\partial X^{2}} - \frac{\partial^{2}}{\partial Y^{2}} + 4r_{s}^{2} e^{2Y} \right) \Psi\left(X,Y\right) = 0,
\end{eqnarray}
where $X = \ln a$ and $Y = \ln b$. Since this equation allows the separation of variables, the generic bounded solution becomes
\begin{eqnarray}
\Psi(X,Y) = \int_{-\infty}^{\infty} f(k) e^{-ikX} K_{ik}\left( 2r_{s} e^{Y} \right) dk,
\end{eqnarray}
where $f(k)$ is an arbitrary mode function of $k$ \cite{Bouhmadi-Lopez:2019kkt}.

The function $f(k)$ is related to the boundary condition of the wave function. The most reasonable way is to assign the classical boundary condition near the horizon, i.e., $X , Y \rightarrow \infty$ limit. For example, if we choose $f(k)e^{ikX} + f(-k)e^{-ikX} = k \sin kX$, then an exact solution becomes \cite{Bouhmadi-Lopez:2019kkt}
\begin{eqnarray}
\Psi(X,Y) = \pi r_{s} e^{Y} \sinh X e^{-2r_{s} e^{Y} \cosh X}.
\end{eqnarray}

\begin{figure}
\begin{center}
\includegraphics[scale=0.5]{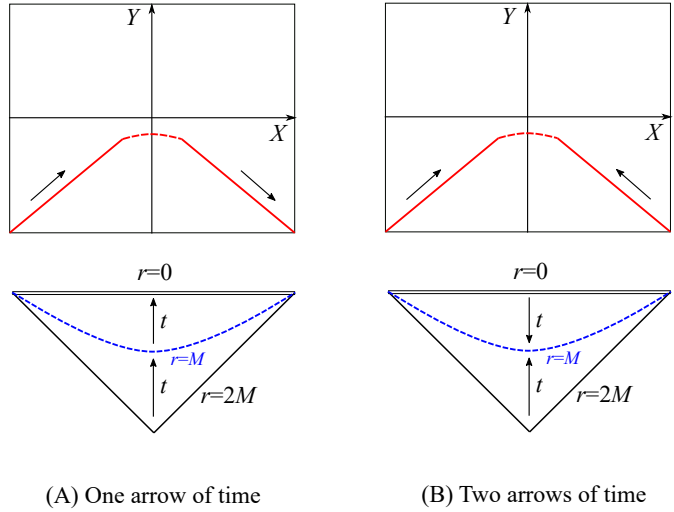}
\caption{\label{fig:pen2}Two interpretations about time \cite{Bouhmadi-Lopez:2019kkt}. (A) there is one arrow of time and (B) there are two arrows of time. The upper figures denote the steepest-descents (red curve) of the wave function on the $X$-$Y$ plane.}
\end{center}
\end{figure}

\section{Two interpretations about time: one arrow or two arrows?}

From the analytic solution, it is easy to confirm that the wave function is proportional to $\tanh X$ along the steepest-descent. Therefore, for the large $|X|$ limit of the steepest-descent, i.e., near the horizon ($X, Y \rightarrow -\infty$) or near the singularity ($X, -Y \rightarrow + \infty$), the probability is almost a constant and hence the steepest-descent can be interpreted as a classical history. However, around the $X = Y = 0$ regime, the wave function is bounced; hence, the $X = Y = 0$ point can be interpreted as a quantum gravitational regime.

In terms of the steepest-descent, two classical domains (near the horizon and near the singularity) must be connected. One way to interpret this is that there is one arrow of time and two classical domains are connected. Then, the steepest-descent says nothing but a classical interpretation of the spacetime; eventually, the universe reaches the singularity and there is no singularity resolution (left of Fig.~\ref{fig:pen2}). The other way to interpret this solution is that two arrows of time are collided at the $X = Y = 0$ point. After the collision, the probability goes to zero and there is no more meaning of the classical spacetime. Therefore, the second interpretation says that two pieces of the classical spacetime are annihilated to nothing (right of Fig.~\ref{fig:pen2}). So, we call this process the \textit{annihilation-to-nothing}.

\section{Connections to the generalized uncertainty principle}

It can be strange that the quantum bouncing happens around the $X = Y = 0$ point, i.e., around the $r \simeq M$ surface. If the black hole mass is large enough, the $r\sim M$ surface is very classical. Then, how can such a classical surface be a place of quantum bouncing or annihilation-to-nothing?

In order to understand this, we invoke the idea of the generalized uncertainty principle (GUP) \cite{Veneziano:1986zf}. The GUP is presented by
\begin{eqnarray}
\Delta x \Delta p \geq  \frac{1}{2} \left( 1 + \alpha \frac{\Delta p^{2}}{M_{\mathrm{P}}^{2}} \right),
\end{eqnarray}
where $\alpha \simeq \mathcal{O}(1)$ is a constant ($c = G = \hbar = 1$ and $G = M_{\mathrm{P}}^{-2}$). If $\alpha \rightarrow 0$, then it returns to the Heisenberg uncertainty principle (HUP).

From GUP, we can derive several useful relations \cite{Chen:2014bva}. For example,
\begin{eqnarray}
\Delta E_{\mathrm{min}} \equiv \frac{M_{\mathrm{P}}^{2}}{\alpha} \Delta x \left( 1 - \sqrt{1 - \frac{\alpha}{M_{\mathrm{P}}^{2} \Delta x^{2}}} \right) \leq \Delta p \leq \frac{M_{\mathrm{P}}^{2}}{\alpha} \Delta x \left( 1 + \sqrt{1 - \frac{\alpha}{M_{\mathrm{P}}^{2} \Delta x^{2}}} \right).
\end{eqnarray}
From this, there are two important points:
\begin{itemize}
\item[--] We require that $\Delta x \geq \Delta x_{\mathrm{min}}$, where $\Delta x_{\mathrm{min}} \equiv \sqrt{\alpha}/M_{\mathrm{P}}$. Therefore, there exists the \textit{minimal length}.
\item[--] The minimum energy to probe $\Delta x$ is $\Delta E_{\mathrm{min}}$.
\end{itemize}

There are two levels of interpretation of GUP around a black hole, where the black hole horizon size is $r_{\mathrm{h}} \sim M \gg l_{\mathrm{P}}$, where $M$ is the mass of the black hole.
\begin{description}
\item[-- Interpretation 1 (\textit{weak version})] the GUP holds for \textit{each independent observer} within a given classical background. Then, for a given black hole background that satisfies the equivalence principle, an observer with energy $\Delta p$ can probe the length scale larger than the uncertainty $\Delta x$:
\begin{eqnarray}
\Delta x \geq  \frac{1}{2\Delta p} \left( 1 + \alpha \frac{\Delta p^{2}}{M_{\mathrm{P}}^{2}} \right).
\end{eqnarray}
Therefore, as long as $M_{\mathrm{P}} \gg \Delta p \gg M_{\mathrm{P}}^{2}/M$, the uncertainty can be smaller than the horizon scale, $\Delta x \ll r_{\mathrm{h}}$, and hence one can probe inside a definite black hole geometry.
\item[-- Interpretation 2 (\textit{strong version})] the GUP holds for the \textit{unitary observer} who can see the entire wave function and hence who lives in the superspace. Then, as the observer probes a place using the energy $\Delta p \sim M$, the length uncertainty becomes
\begin{eqnarray}
\Delta x \geq  \frac{1}{2 M} \left( 1 + \alpha \frac{M^{2}}{M_{\mathrm{P}}^{2}} \right) \simeq M.
\end{eqnarray}
Therefore, no unitary observer can probe smaller length scale than $\Delta x \simeq M$.
\end{description}

There is a tension between two interpretations. When we study thermodynamics from the GUP, we regard that $\Delta x \simeq M$ and calculate the minimal energy $\Delta p$ as the Hawking temperature:
\begin{eqnarray}
T \propto \frac{M_{\mathrm{P}}^{2}}{\alpha} \Delta x \left( 1 - \sqrt{1 - \frac{\alpha}{M_{\mathrm{P}}^{2} \Delta x^{2}}} \right) \simeq \frac{1}{M}.
\end{eqnarray}
Hence, for thermodynamical applications of the GUP, we always rely on this second interpretation. On the other hand, it is very strange to accept that the uncertainty radius is extremely large and hence impossible to probe inside the black hole.

The tension between the two interpretations is deeply related to the problem of the measurement of quantum mechanics. After the measurement, one can see a classical observable which is no more unitary. If we carefully consider the difference between two observers (unitary observer and semi-classical observer), then one can reconcile the tension about two interpretations. In other words, we can think that the weak interpretation is obtained after the \textit{Everett branching}, while the strong version is true for the entire wave function (see Table 1).

\begin{table}[t]
\begin{center}
\begin{tabular}{c|c|c|c}
\hline \hline
\multicolumn{2}{c|}{\;\;\;Quantum mechanics (HUP)\;\;\;} & \multicolumn{2}{|c}{\;\;\;Quantum gravity (GUP)\;\;\;}\\
\hline
\begin{tabular}{c}
\;\;\;\;\;\;Unitary observer\;\;\;\;\;\;\\
(in space)
\end{tabular} &
\begin{tabular}{c}
$~$\\
\;\;\;\;\;\;After measurement\;\;\;\;\;\;
\\$~$
\end{tabular} &
\begin{tabular}{c}
\;\;\;\;\;\;Unitary observer\;\;\;\;\;\;\\
(in superspace)
\end{tabular} &
\begin{tabular}{c}
\;\;\;\;\;\;After Everett branching\;\;\;\;\;\;
\end{tabular}\\
\hline \hline
\begin{tabular}{c}
$\Delta x \geq \frac{1}{2\Delta p}$,\\
where $\Delta p > 0$\\
is the energy that the\\
unitary observer used.
\end{tabular}
& \begin{tabular}{c}
$\Delta x = 0$ or $\Delta p = 0$,\\
and possible to probe\\
a definite position.
\end{tabular}
& \begin{tabular}{c}$~$\\
$\Delta x \simeq M \simeq r_{\mathrm{h}} > 0$,\\
where $\Delta p \simeq M \gg M_{\mathrm{P}}$\\
is the energy that the\\
unitary observer used.\\
$~$\\
\end{tabular}
& \begin{tabular}{c}
$\Delta x \geq \frac{1}{2\Delta p} \left( 1 + \alpha \frac{\Delta p^{2}}{M_{\mathrm{P}}^{2}} \right) $,\\
where $M_{\mathrm{P}} \gg \Delta p \gg M_{\mathrm{P}}^{2}/M$\\
is the energy that the\\
free-falling observer used.\\
Hence, $\Delta x \ll r_{\mathrm{h}} \simeq M$ and\\
possible to probe\\
inside the horizon. 
\end{tabular}\\
\hline
\begin{tabular}{c}
Observables $\mathcal{O}$ are\\
approximated by\\
expectation values, e.g.,\\
$\langle \mathcal{O} \rangle \simeq \sum p_{i}\mathcal{O}^{(i)}$.
\end{tabular}
& \begin{tabular}{c}
Has a definite observable\\
eigen value,\\
e.g., $\mathcal{O}^{(i)}$.
\end{tabular}
& \begin{tabular}{c}
Observables $\mathcal{O}$ are\\
approximated by\\
expectation values, e.g.,\\
$\langle g_{\mu\nu} \rangle$ or $\langle \phi \rangle$.
\end{tabular}
& \begin{tabular}{c}
$~$\\
Has an approximately definite\\
observable, e.g., $g_{\mu\nu}^{(i)}$ or $\phi^{(i)}$\\
even inside the horizon scale.
\\$~$
\end{tabular}\\
\hline \hline
\end{tabular}
\caption{\label{tab:ana}Analogy between quantum mechanics (with HUP) and quantum gravity (with GUP).}
\end{center}
\end{table}

\begin{figure}[h]
\begin{center}
\includegraphics[scale=0.3]{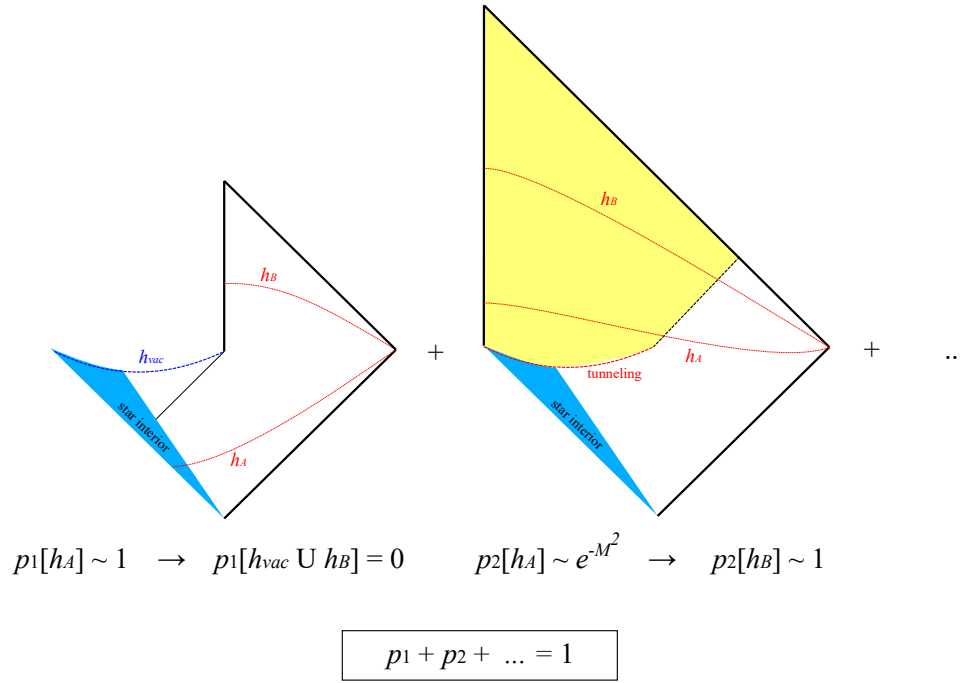}
\caption{\label{fig:pen5}The first one is the most probable semi-classical history, but due to the annihilation-to-nothing surface $h_{vac}$, the probability will decay to zero. Hence, eventually, the trivial geometries will be dominated (see detailed explanations in \cite{Bouhmadi-Lopez:2019kkt}).}
\end{center}
\end{figure}

\section{Revisit the resolution of the information loss problem}

The two-fold nature of GUP between two observers (unitary observer and semi-classical observer after the Everett branching) does indeed appear if we consider the entire wave function to understand the information loss problem \cite{Sasaki:2014spa} (Fig.~\ref{fig:pen5} and \cite{Bouhmadi-Lopez:2019kkt}).

The wave function will be approximated by a superposition of semi-classical histories. For the most probable history, a black hole will be formed, evaporated, and disappeared, but its inside should be treated by the annihilation-to-nothing hypersurface. Therefore, the probability of the history will eventually decay to zero, although outside the horizon can still remain a semi-classical geometry.

On the other hand, there also exists an exponentially suppressed history \cite{Chen:2018aij}; due to the quantum tunneling, there is no formation of a singularity nor an event horizon. For this history, the information will be preserved, but the probability weight can be negligible \cite{Maldacena:2001kr}. However, if the probability of the most probable history eventually decreases because of the annihilation-to-nothing surface, then the other histories without a singularity will be eventually dominated (if the sum of all probabilities must be preserved). This results that even though the history with a trivial geometry has an exponentially suppressed probability, this will be dominated; information will be preserved via these trivial geometries \cite{Sasaki:2014spa}.

If we superpose all geometries, then due to the probability changing nature, the uncertainty must be dominated around the horizon scale. This horizon scale uncertainty is also consistent with the nature of the strong interpretation of GUP. Such uncertainty can be only seen by a unitary observer; if the observer is semi-classical, then there is no superposition and hence there is no $\Delta x \sim M$ scale uncertainty.

\section{Conclusion}

In this paper, we briefly reviewed the paper \cite{Bouhmadi-Lopez:2019kkt} and investigated quantum gravitational wave function inside a black hole. Due to the ambiguity of the interpretation of time, one can introduce two arrows of time. This implies that two classical spacetime seems to be annihilated to nothing. This new interpretation gives a clue to understanding the information loss problem.

Since the annihilation-to-nothing happens around a very large length scale (still inside the horizon), one can ask how can it happen in reality. One potential answer is to apply for the generalized uncertainty principle. It is very reasonable that the GUP implies the large-length scale uncertainty, at least for unitary observers. Usual semi-classical observers will lose such a large-length scale uncertainty after the Everett branching. Analogously, two observers can be introduced in order to see the entire wave function; then the unitary observer will see large-length scale uncertainty and lose the semi-classical geometry, while the semi-classical observer will see the equivalence principle but lose the unitarity.

This new observation and interpretation will provide a consistent view of the information loss paradox. We believe that this will shed some light on the ultimate and consistent understanding of quantum gravity as well.

\begin{acknowledgments}
The author would like to thank Mariam Bouhmadi-Lopez, Suddhasattwa Brahma, Che-Yu Chen, Pisin Chen, and Misao Sasaki for stimulating discussions and comments. This work was supported by a 2-Year Research Grant of Pusan National University.
\end{acknowledgments}

\end{document}